\documentclass[aps,preprint,showkeys,
superscriptaddress,
amsmath,amssymb,
aps,
]{revtex4-1}

\usepackage{graphicx}% Include figure files
\usepackage{dcolumn}% Align table columns on decimal point
\usepackage{bm}% bold math
\usepackage{color}
\usepackage[utf8]{inputenc}

\usepackage{hyperref}

\usepackage{mathtools}
\usepackage{empheq}
\usepackage[skins,theorems]{tcolorbox} %boxes setting
\tcbset{highlight math style={enhanced,
  colframe=black,colback=white,arc=4pt,boxrule=1pt}}

\usepackage{algpseudocode}
\usepackage{algorithm}
\usepackage[english]{babel}

\usepackage{etoolbox}
\usepackage{amsmath}

\newcommand{\algstrut}[1][\algruledefaultfactor]{\vrule width 0pt
depth .25\baselineskip height #1\baselineskip\relax}
\newcommand*{\algrule}[1][\algorithmicindent]{\hspace*{0.5em}\vrule\algstrut
\hspace*{\dimexpr#1+1.0em}}%aumenta disminuye distancia horizontal indexado for

\makeatletter
\newcount\ALG@printindent@tempcnta
\def\ALG@printindent{%
    \ifnum \theALG@nested>0% is there anything to print
    \ifx\ALG@text\ALG@x@notext% is this an end group without any text?
    \else
    \unskip
    \ALG@printindent@tempcnta=1
    \loop
    \algrule[\csname ALG@ind@\the\ALG@printindent@tempcnta\endcsname]%
    \advance \ALG@printindent@tempcnta 1
    \ifnum \ALG@printindent@tempcnta<\numexpr\theALG@nested+1\relax% can't do <=, so add one to RHS and use < instead
    \repeat
    \fi
    \fi
}%

\patchcmd{\ALG@doentity}{\noindent\hskip\ALG@tlm}{\ALG@printindent}{}{\errmessage{failed to patch}}

\AtBeginEnvironment{algorithmic}{\lineskip0pt}

\begin{document}

\date{\today}

%\preprint{APS/123-QED}

\title{Artificial Intelligence, Chaos, Prediction and Understanding in Science}% Force line breaks with \\

\author{Miguel A. F. Sanju\'{a}n}
\email{Corresponding Author : miguel.sanjuan@urjc.es}
\affiliation{Nonlinear Dynamics, Chaos and Complex Systems Group, Departamento de  F\'isica, Universidad Rey Juan Carlos, M\'ostoles, Madrid, Tulip\'an s/n, 28933, Spain}
\affiliation{Department of Applied Informatics, Kaunas University of Technology, Studentu 50-415, Kaunas LT-51368, Lithuania}

\begin{abstract}

Machine learning and deep learning techniques are contributing much to the advancement of science. Their powerful predictive capabilities appear in numerous disciplines, including chaotic dynamics, but they miss understanding. The main thesis here is that prediction and understanding are two very different and important ideas that should guide us about the progress of science. Furthermore, it is emphasized the important role played by nonlinear dynamical systems for the process of understanding. The path of the future of science will be marked by a constructive dialogue between big data and big theory, without which we cannot understand.

\end{abstract}
%%%%%%%%%%%%%%%%%%%%%%%%%%%

%%%%%%%%%% Insert the texts which can accomdate on firstpage in the tag "fmtext" %%%%%
\maketitle

\section{Introduction}

Techniques from artificial intelligence are contributing to transform every walk of life, enabling people to analyze data, integrating information and providing ways to improve decision-making, and fields such as the biotechnology, health care, speech and voice recognition, transport, finance, and climate change, among others. Computers learn from numerous examples, after having been trained, and can recognize patterns in big data sets, and classify them in different categories. As a consequence, they are able to carry out certain specific tasks with a much higher precision than humans. Many criticize that this does not signify that the machine is intelligent, since intelligence is something far more complex, and furthermore, most of the serious scientists agree that we are extremely far away from a machine being more intelligent than a human being. No doubt, it would be fantastic that AI would be more versatile, since right now in nearly all cases everything that has been achieved is related to pattern recognition, when practically all interesting problems are definitely much more complicated than that.

One of the main threads of my article will be the analysis between artificial intelligence and its relationship between prediction and understanding in science. Another key idea I wish to emphasize throughout this article is the important role that ideas from nonlinear dynamics and dynamical systems theory play in the process of understanding science, and the evolutionary dynamics of physical and biological processes. As it will be commented later, even neuroscientists associate the very meaning of understanding to dynamical systems theory.

At the heart of the scientific endeavour lies the desire of understanding the universe, knowing what kind of reasons explain the past events, and acquiring the ability of forecasting the future. Since the earliest times the main task of a scientist today is to observe nature, to build models from the observations, and to use them for predictions. Forecasting is the process of making predictions of the future based on past and present data. Thanks to the scientific models, it is possible to understand nature and the mechanisms that explain the observations, attempt to forecast extreme events and the weather, to prevent diseases, calculate the position of the celestial bodies, as well as develop the astronautic technologies, understand the behavior of the components of matter, fight against epidemics, etc.

Nevertheless, the importance of forecasting goes beyond the practical purposes and points to the essence of the scientific method itself. When we analyze a certain scientific problem, one of the goals is to catch the reality as faithful as possible from a model, with the expectation to obtain an appropriate understanding of the involved physical processes. That is why making excellent predictions with our model and to test the predictions with new observations is so important for the development of science by using the scientific method, as well as for our true understanding of the universe.

Since the beginning of science, there has been an stimulating interaction between science and philosophy, though this relationship has not always received the same interest from both parts in the last decades. Actually, the English word \textit{scientist} was first coined by William Whewell in 1834. And is well known that before that, the name used was \textit{natural philosopher}. I will begin by commenting on the need that science has of philosophy, as a group of scientists analyze in a very recent article published in the prestigious journal Proceedings of the National Academy of Sciences of the United States of America, \cite{laplane_2019} as well as other recent references that talk about the relationship between physics and philosophy \cite{rovelli_2018}, and physics and history \cite{stanley_2016}.

The authors of \cite{laplane_2019} argue that philosophy can have an important and productive impact on science, and provide a series of recommendations to create a better atmosphere and dialogue between science and philosophy. After a persuasive discussion on the positive aspects of this dialogue, they basically conclude that: ``Modern science without philosophy will run up against a wall: the deluge of data within each field will make interpretation more and more difficult". Something that definitely is of the most importance considering our era of big data.

On similar grounds the physicist Carlo Rovelli in his essay \textit{Physics Needs Philosophy. Philosophy Needs Physics} \cite{rovelli_2018} argues in defense of the influence on physics that philosophy has had, as well as the influence of physics in philosophy. The emphasis is mostly done on the constructive role to conceptualize through theories after a simple recollection of data, also of much interest in our discussion here.

Another thought-provoking article in this context has been written by Matthew Stanley with the title \textit{Why should physicists study history?} \cite{stanley_2016}, where he emphasizes the utility of knowing the historical aspects and social interactions that affect the evolution of physics. Furthermore, it provides an intellectual flexibility exposing scientists to new ways of thinking and forcing them to reexamine what is already known. Certainly, a knowledge of history can help us have an enriching reflection on how we know what we know and how it could be otherwise.

Truly in recent times there have been spectacular developments made by machine learning and deep learning techniques in relation to numerous scientific predictions. These include chaotic systems as well, where is well known that they have prediction problems.

Among them, we can highlight the tremendous impact of AlphaGo Zero \cite{silver_2016} and AlphaZero \cite{silver_2018} that defeated the world’s best Go players and the best chess computer programs, respectively. The fascinating thing about these programs is that they are able to perform very specific and well defined tasks in a extraordinarily well manner. However, the programs do not analyze the plays the way humans do. It happens that even the same programmers who write the computer code do not understand why the programs make certain decisions.
 
\begin{figure}

\begin{center}
\includegraphics[width = 9cm]{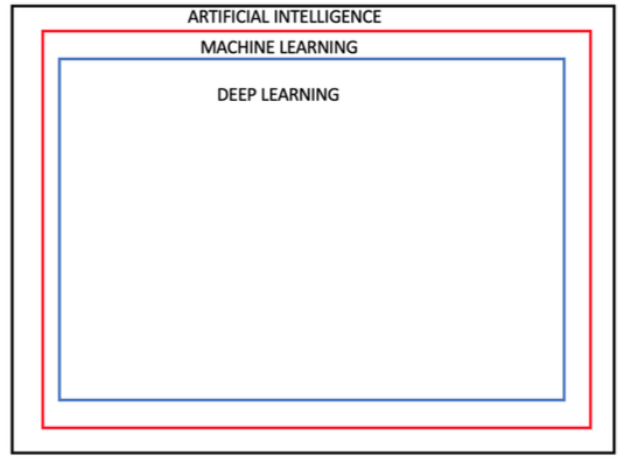} 
\end{center}

\caption{
Artificial intelligence, machine learning and deep learning occupies almost the same space nowadays, when in the recent past deep learning was simply a small part of machine learning and machine learning a small part of AI.
}
\label{ai_mldl}
\end{figure}

Interestingly, in the recently published book \textit{Artificial Intelligence: A Guide for Thinking Humans} \cite{mitchell_2019} Melanie Mitchell discusses, among other captivating issues, the evolution of the artificial intelligence methods during the last decades. And she describes that artificial intelligence, machine learning and deep learning occupies almost the same space nowadays, when in the recent past deep learning was simply a small part of machine learning and machine learning a small part of AI (See Fig.~\ref{ai_mldl}).

%%%%%%%%%%%%%%%%%%%%%%%%%%%%%%%%%%%%

Enthusiasm is a necessary step to go ahead in any human enterprise, but it is also wise to see whether some claims are real and true, since enthusiasm sometimes has replaced cool heads. Artificial intelligence has generated, needless to say, much enthusiasm, and this has also provoked certain reactions pointing out the flaws of some of the extreme claims, that will be discussed later on in this article. Naturally, in spite of all the enthusiasts on AI, there are certainly critics. In particular, Artificial intelligence owes a lot of its smarts to Judea Pearl, who won the Turing Award in 2011. In the 1980s he led efforts that allowed machines to reason probabilistically. But, now he is one of the field’s sharpest critics. In his latest book, \textit{The Book of Why: The New Science of Cause and Effect}, \cite{pearl_2018} he argues that artificial intelligence has been handicapped by an incomplete understanding of what intelligence really is. He has also declared recently that ``All the impressive achievements of deep learning amount to just curve fitting,'' \cite{hartnett_2018}. He also defends the idea to teach machines to understand why questions, by basically replacing reasoning by association with causal reasoning. This certainly goes to the core question of understanding.

The paper is structured as follows. Section~2 is devoted to a general discussion on chaos and predictability, including recent developments on chaos and machine learning and  the predictability derived from the presence of fractal structures in phase space. A further discussion on hetero-chaos, UDV and prediction, including shadowing will be discussed in Sect. ~3. Section~4 is focused in giving examples about the differences between the two different notions of prediction and understanding. In Sect.~5 different ways of understanding are commented, as well as understanding by machines. Section~6 describes how recent developments of machine learning has brought some authors to deny the value of the scientific method, and the reaction of many scientists to this situation. The paper ends emphasizing the conclusions on the importance of keeping the prediction and understanding close together and claiming for a constructive dialogue between data-driven models and theoretical and conceptual models, as well as the important role that dynamical systems play for understanding.

\section{Chaos and predictability}

\subsection{Chaos and machine learning}

Chaos theory has shown that long-term prediction is impossible. The slightest disturbance of a chaotic system can lead us to be unable to specify the future state with sufficient precision so that we cannot predict its evolution, what implies an intrinsic situation of uncertainty.
In recent work by Ed Ott and collaborators \cite{pathak_2017,pathak_2018a} having used machine learning techniques, they have reported to be able to predict the future evolution of chaotic systems with further precision than before, by extending the future horizon of the prediction further ahead to what it could be done with current algorithms. 

They employed a machine-learning algorithm called reservoir computing to learn the dynamics of a well-known spacetime chaotic dynamical system used to study turbulence and spatiotemporal chaos, called the Kuramoto-Sivashinsky equation. The important result lies at the fact that after training the equation with past data, they were able to predict the evolution of the system out to eight Lyapunov times into the future, what basically means eight times further ahead than previous methods allowed. The Lyapunov time represents how long it takes for two almost-identical states of a chaotic system to exponentially diverge. It represents the inverse of the largest Lyapunov exponent of a dynamical system. As such, it typically sets the horizon of predictability. The algorithm knows nothing about the dynamical system itself; it only sees data recorded about its evolving solution. In essence, the results suggest that you can make the predictions with only data, without actually knowing the equations. 

In another research published in \cite{pathak_2018b} by the same group, they showed that improved predictions of chaotic systems like the Kuramoto-Sivashinsky equation become possible by hybridizing the data-driven, machine-learning approach and traditional model-based prediction, so that accurate predictions have been extended out to twelve Lyapunov times, what suggests the importance of integrating both methods.

A discussion on the relation on data science and dynamical systems theory is given in \cite{berry_2020}. The authors combine ideas from dynamical systems theory and from learning theory as a way to create a more effective framework to data-driven models for complex systems. They clearly comment that in spite of the tremendous successes of statistical models of complex systems, these models are treated as black-boxes with limited insights about the physics involved and lacking understanding. They describe mathematical techniques for statistical prediction phenomena usually studied in nonlinear dynamics. In spite of the numerous mathematical techniques reviewed at the interface of dynamical systems theory and data science  for  statistical  modeling  of  dynamical  systems, they do not discuss recent developments in deep learning or reservoir computing.

A notorious impact has received in a similar context a recent paper \cite{breen_2019}, where the main goal has been to solve the chaotic three-body problem using deep neural networks. The main idea is the use by the authors of an integrator for an n-body problem focusing in the three-body problem. The data they obtain with the integrations are used to train a neural network so that they are able to obtain and predict trajectories much ahead the previous predictions, and in a very fast manner. The three-body problem is one of the classical unsolved problems in physics that was formulated by Newton, which basically consists on solving the equations of motion for three bodies under their own gravitational force. This constitutes also a classical example of chaos in Physics after Poincaré proved its non-integrability and its chaotic nature already at the end of the 19th century \cite{poincare_1890,poincare_1892}. The authors show that an ensemble
of solutions obtained using an arbitrarily precise numerical integrator can be used to
train a deep artificial neural network that, over a bounded time interval, provides accurate solutions at fixed computational cost and up to 100 million times faster than a state-of-the-art solver. The main applications they have in mind are in astrophysics, black-hole systems or galactic dynamics. The success in accurately reproducing the results of the three-body problem, a classical chaotic system, provides an stimulus for solving other chaotic problems of similar complexity, by basically substituting classical solvers with machine learning algorithms trained on the underlying physical processes \cite{pathak_2018a,stinis_2019}.

\subsection{Predictability, attractors and basins }

Issues related with chaos and prediction are very important in science and much discussed by many authors in the context not only of dynamical systems but in relation to different scientific disciplines. In Physics we have laws that determine the time evolution of a given physical system commonly modeled by a dynamical system, depending on its parameters and its initial conditions. Precisely with these laws we can predict the future evolution of the phenomena we model. However, not always a good prediction can be done. Oftentimes and in particular when we have nonlinear or chaotic systems predictions are limited, what implies an intrinsic unpredictability. There are different sources of uncertainty and unpredictability in dynamical systems. A small uncertainty in the initial conditions gives rise to a certain unpredictability of the final state as a consequence of the sensitive dependence on initials conditions, which is a typical hallmark of chaos. Another source of uncertainty are the fractal structures commonly appearing in the basins of attraction in phase space. Chaotic systems typically present fractal basins. 

Given a dynamical system possessing only one attractor in a certain region of phase space, then the final state of the system is uniquely determined for any initial condition. Nevertheless, in most cases dynamical systems may possess more than one attractor in the same region of phase space, that is, the system is multi-stable, so that to elucidate which orbits tend to which attractor becomes a key issue. In multi-stable systems with many basins of attraction, the dynamical system may possess fractal or even Wada boundaries so that the prediction becomes harder, fundamentally based on the uncertainty associated to the initial conditions. A thorough review of fractal basins and fractal structures in nonlinear dynamics can be found in \cite{aguirre_2009}.

Much work has been made in the past few years to clarify different aspects of unpredictability in dynamical systems \cite{vallejo_2019} \cite{saiki_2018}. Among other efforts, the new notion of basin entropy \cite{daza_2016} provides a new quantitative way to measure the unpredictability of the final states by analyzing basins of attraction.  A detailed discussion of the issue of predictability and basins of attraction appears in the book \cite{vallejo_2019}.

\section{Hetero-Chaos, UDV and prediction}
\label{hc}
\subsection{Predictability and shadowing}

\begin{figure}

\begin{center}
\includegraphics[width = 9cm]{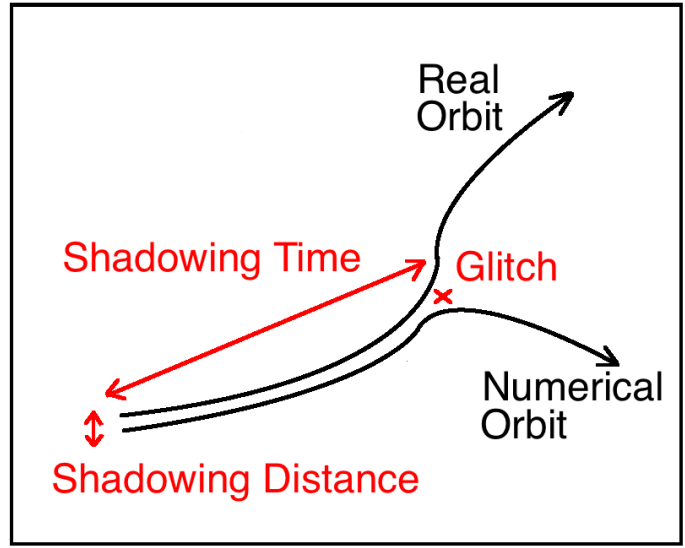} 
\end{center}

\caption{
The shadowing time is the time a numerical trajectory remains close to a true trajectory which is called a \emph{shadow}. The distance to the shadow might be seen as an observational error, within which the computer-generated orbits are considered reliable. At the glitch, the true trajectory diverges from the computed-generated trajectory.
}
\label{shadow}
\end{figure}

We are typically used to associate chaos with a lack of predictability. However, this is not always the case. As a matter of fact, any scientist is commonly faced with the key question of knowing how good a numerical prediction of a model is, and for how long is valid. Quantitative answers to these questions were given by \cite{sauer_1997} by using the concepts of shadowing distance, that measures the distance from the shadowing trajectory to the computer-generated trajectory, and shadowing time, that measures the length of true shadowing trajectories. This shadowing time will be the basis to assess the predictability of our models. This idea is illustrated in Fig.~\ref{shadow}.

The shadowing property is frequently of great interest in practice. Often, one wants to know whether there exists a true orbit that closely follows a given computer-generated orbit, where the noise results from roundoff error. A computer simulation of a chaotic system can be distinct from the true trajectory. It may happen that after a certain time interval the distance between the true orbit and the computed orbit increases greatly. Nevertheless, it is possible that a true orbit might shadow the computed trajectory what means that it remains close to it for a long computing time.

%%%%%%%%%%%%%%%%%%%%%%%%%%%%%%%%%%%%% FIGURE 5
%%%%%%%%%%%%%%%%%%%%%%%%%%%%%%%%%%%%

The shadowing time is directly linked to the hyperbolic or nonhyperbolic nature
of the orbits. For hyperbolic chaotic systems, where the angle between the stable and unstable manifolds is away from zero and the phase space is locally spanned by a fixed number independent stable and unstable directions, the shadowing is present during long times and numerical trajectories stay close to the true ones. Even though there might be some exceptions, it is rather common to accept that for most dynamical systems the shadowing property and hyperbolicity are indeed equivalent.

Nevertheless, most chaotic attractors of physical interest are not hyperbolic. In most cases, the attractors fail to be hyperbolic due to homoclinic tangencies, where stable and unstable manifolds intersect tangentially. Another mechanism for the appearance of nonhyperbolicity is due to the presence of Unstable Dimension Variability (UDV) \cite{kostelich_1997}, where the dimension of the unstable and stable tangent spaces is not constant. In these cases, an orbit may be shadowed, but only for a very short time, and the computed orbit behavior may be completely different from the true one after this period of time.

\subsection{Hetero-chaos}

Some of these previous issues have been recently discussed in the context of the new concept of hetero-chaos \cite{saiki_2018}. The presence of hetero-chaos has serious consequences for the predictability of chaotic systems, that are common in science. As a matter of fact, predictability is more difficult when a chaotic attractor has different regions that are unstable in more directions than in others. This means that arbitrarily close to each point of the attractor there are different periodic points with different unstable dimensions. When this happens, we say the chaos is heterogeneous, in contrast to homogeneous chaos occurring when there is only one unstable dimension, and the phenomenon receives the name of hetero-chaos.

A relevant issue to our previous discussion on prediction and shadowing is also derived from hetero-chaos. As it was mentioned earlier, knowing how good a numerical simulation works and for how long the computed orbit is valid is of the most importance for a scientist using numerical simulations of a model. The shadowing property makes that our simulations are realistic, but they become unrealistic when shadowing fails, that may occur when the number of unstable directions increases for a trajectory in phase space. This transition from a lower to a higher number of unstable directions has dynamical consequences that are manifested through the fluctuations around zero of the finite time Lyapunov exponents, something typically happening for higher-dimensional dynamical systems. This is also a common mechanism for the appearance of nonhyperbolicity, and as a consequence shadowing fails \cite{dawson_1994}. This poses a serious difficulty for predictability since hetero-chaotic systems cannot have the shadowing property while homogeneous chaotic systems usually do have. 

A short comment on Unstable Dimension Variability (UDV), that occurs when an attractor has two periodic orbits that are unstable in different numbers of dimensions. As a consequence a typical trajectory on the chaotic attractor will visit small neighborhoods of saddles and repellers in the attractor, so that a common good indicator of the UDV is precisely the fluctuations about zero of the finite time Lyapunov exponents. 

As the authors of \cite{saiki_2018} express, hetero-chaos means that unstable periodic orbits embedded in a chaotic set have distinct numbers of unstable directions. Accordingly, a  trajectory will typically  move  in  regions  with different  unstable  dimensions,  leading  to  fluctuations  about zero  of  some  Lyapunov  exponents, and affecting the shadowing property and its predictability. From this point of view, it can be understood as a unifying concept comprising different phenomena observed in numerical simulations of chaotic dynamical systems and physical experiments, such as UDV, on-off intermittency, riddled basins, blowout and bubbling bifurcations, where common patterns are present. Moreover, they conjectured that UDV almost always implies hetero-chaos. Since shadowing fails for hetero-chaotic systems, ascertaining when a homogeneous chaotic system becomes heterogeneous is paramount when we are discussing predictability. In addition, considering that models with high dimensional chaotic attractors are receiving much more attention by many researchers as models of numerous physical phenomena, this indicates the relevance of this issue as what concerns prediction of physical systems. Hetero-chaos seems to be important for most physical systems with high-dimensional attractors, including weather prediction and climate modelling, what also shows a serious limitation to predictability either achieved with ordinary methods or methods derived from artificial intelligence.

\section{Prediction and understanding}

\begin{figure}
\begin{center}
\includegraphics[width=0.7 \textwidth]{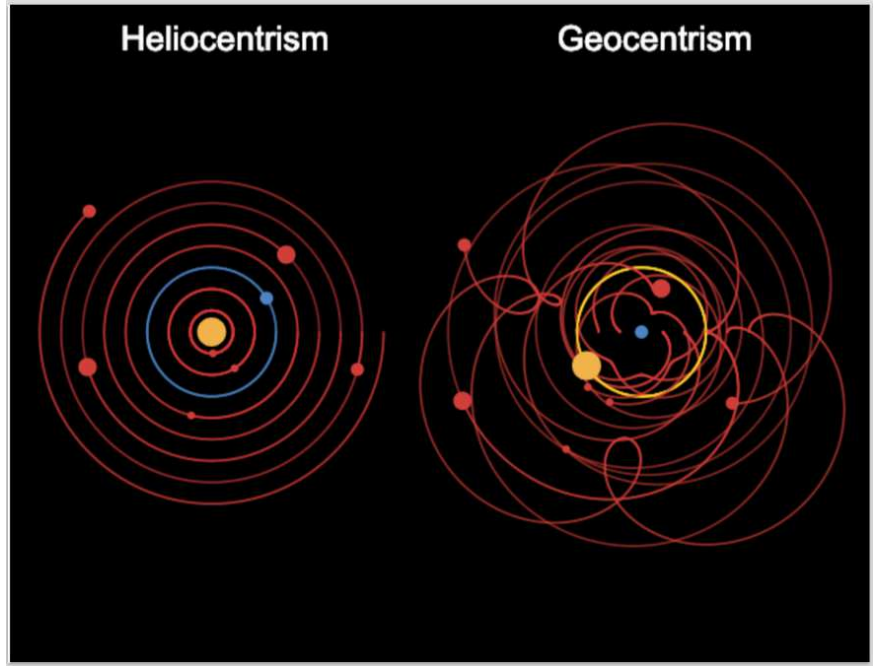}
\caption{The figure shows the trajectories of the planets of the solar system by using the geocentric and the heliocentric model. (Taken from \cite{christersson})
\label{geo_helio}
}% end of caption
\end{center}
\end{figure}

When we approach the issue of machine learning and understanding in science, important questions arise. As a matter of fact, an excellent ability in prediction could not imply a correct understanding of the physical processes involved. Prediction and understanding are certainly two different concepts. Actually, to illustrate this idea we can draw inspiration from the history of the planetary motion. We can start with Ptolemy's methods and his geocentric method to predict how planets move in the sky. As is well known, Ptolemy did not know the theory of gravity, not even that the sun occupied the center of the solar system. While it was possible to predict the motions of the planets, it was not known why these methods worked. This theory lasted for a quite long time and it was followed by the work of several brilliant scientists. Years later the heliocentric system of Nicolaus Copernicus changed everything. This is illustrated in the Fig.~\ref{geo_helio}. Later at the dawn of the modern times came the astronomical observations of Galileo Galilei, that were continued by the work of Johannes Kepler and his famous laws.

And at the end, Isaac Newton found the differential equations that governed the motion of the planets. This was a highly important step, since that contributed to understand why the planets move. The Universal Law of Gravitation formulated in 1687 \cite{newton_1687}, allowed to successfully explain the motion of the planets, from Mercury up to Saturn, already known from ancient times. The same idea, that of finding the differential equation, is the key to understanding, and as a consequence to predict even the existence of other planets.

That was the case of the planet Uranus that was discovered by the British astronomer Frederick William Herschel in 1781.  Once it was realised that it was a genuine planet, further observations continued the following decades that revealed substantial deviations from the tables based on predictions done by the Newton's law of universal gravitation. So confident was the scientific community in the goodness of the Newton's laws, that it was hypothesised that an unknown body was perturbing the orbit through gravitational interaction. The position of this body was predicted in 1846 by the French mathematician Urbain Le Verrier and finally the planet Neptune was found. Some years later, and after careful analyses of its orbit the existence of another new planet was predicted, leading to the discovery of Pluto by the American astronomer Clyde Tombaugh in 1930. All these successes gave strong confidence in the infallibility of the Newton's law of universal gravitation. Even it was postulated the existence of the planet Vulcan \cite{levenson_2015} between the Sun and Mercury, that would explain the perihelion precession of Mercury, but in this case the problem was solved by changing the Newton gravitational law by the Einstein General Relativity theory.

In science the notion of understanding leads us to a similar pattern. Reducing a complicated phenomenon to a simple set of rules or principles, implies an understanding of the considered phenomenon. Machines make their predictions much better than us. But they are not able to explain why. Certainly, artificial intelligence techniques are contributing and will contribute much in science. The predictions can be excellent, but the key issue is whether we can understand what is happening. Prediction without understanding affects the very notion and sense of scientific knowledge as we know it today. Needless to say, there are innumerable unknowns, and all this discussion is not simple at all. However, the important issue is to elucidate the authentic meaning of science. We understand science as the ability to know, understand and predict. Keeping only the predictive capacity is not enough. If we forget understanding, then we could conclude that machines could successfully develop scientific work by themselves.

This tension between prediction and understanding has been permanent in the history of science, as is the case in fields where there exist a large amount of data such as genomics, computational biology, economy and finance. What is usually missing is understanding. But not always this tension has been derived by data. As an example, I will mention a discussion made by Alex Broadbent in his article \textit{Prediction, Understanding, and Medicine} \cite{broadbent_2018}, where he argues that understanding is the core intellectual competence of medicine and as a practical consequence comes the ability to make predictions about health and disease.

There are different ways of doing science, or characteristics and aspects of science that are more relevant in some disciplines than others. The following characteristics might help to classify different scientific disciplines, though in some sense all of them might be necessary.

\begin{itemize}
\item Understanding\\
This attempts mainly to the formulation of important questions in science. Physics is one of them, where after relevant questions we expect to have the answers to the why of things. But of course, the same pattern affects to natural sciences whenever we ask deep questions that we want to answer. Do neutrinos have mass? And if so, why? Why we sleep? Why stars shine? As an answer of these questions we genuinely look for a clear understanding of how things work the way they do.

\item Prediction\\
This is another key aspect of the scientific endeavour. We want to know what will happen. According to what we know we want to predict something unknown. This characteristic is so fundamental, that even it could be argued that if you cannot truly predict a phenomenon you cannot consider it under a scientific discipline. And consequently if you cannot predict you cannot understand. We can predict solar cycles, failures in engineering designs, and natural disasters. The predictive power of science is one of the driving forces of progress and development.

\item Description\\
Clearly, this is another important aspect of science that not necessarily needs logical deductions of the same nature as the why questions. It concerns mainly with answering what and how questions. There are certain scientific disciplines where this characteristic is more common than others. What is consciousness? How did life begin? How a Lyapunov exponent evolve with time? Or merely consider a taxonomy of some concepts or natural objects, a mere description of natural phenomena without going any further.
\end{itemize} 

We can learn physics and predict in physics or other sciences through machine learning, but we still do not know if machines can actually understand. Actually, according to some philosophers and neuroscientists we do not even know what it means to understand. There is another issue that we should discuss here. AI is not able to make interpretations. Unquestionably, they are highly sophisticated optimization algorithms that constantly feed on data until they find enough patterns to make their own predictions. Nevertheless, these patterns are purely empirical laws; they have no theoretical basis or physical interpretation, such as Kepler's or Maxwell's laws. 

Precisely in this context is worth to mention again the critical view of certain developments of AI that are mainly based on data done by Judea Pearl in \cite{pearl_2018}, where he affirms: “In certain circles there is an almost religious faith that we can find the answers to these questions in the data itself, if only we are sufficiently clever at data mining. However, readers of this book will know that this hype is likely to be misguided. The questions I have just asked are all causal, and causal questions can never be answered from data alone. They require us to formulate a model of the process that generates the data, or at least some aspects of that process. Anytime you see a paper or a study that analyzes the data in a model-free way, you can be certain that the output of the study will merely summarize, and perhaps transform, but not interpret the data”.

\section{Different forms of understanding}

We can learn and predict in science through machine learning, but we still do not know if it can be understood. Then, a key question arises: What does understanding mean?. This is precisely the question that the neuroscientist Gilles Laurent \cite{laurent_2000} raises himself in a short essay, where he highlights the importance of the power of explanation of theory, since in order to understand the brain it is necessary to understand a system of interacting elements, the neurons, and how their interactions and structure generate functions. Actually, he strongly emphasizes the role played by the theory of dynamical systems that definitely contribute to help us to have a mental and mathematical conceptualization, and ultimately to understand.

As a matter of fact, even though philosophers have been worried about the meaning of understanding, it seems that it is not something very clear according to the philosopher R. L. Franklin in his article \textit{On Understanding} \cite{franklin_1983}, when he dares to write: ````Understand" is a word we understand as well as any, but we do not understand philosophically what it is to understand."

In any case, in his discussion on the subject, he points out that the notion of understanding is linked to the capacity to explain something, and the explanation is often linked to the causal law that relates what we observe.

\begin{figure}
\begin{center}
\includegraphics[width=1.0 \textwidth]{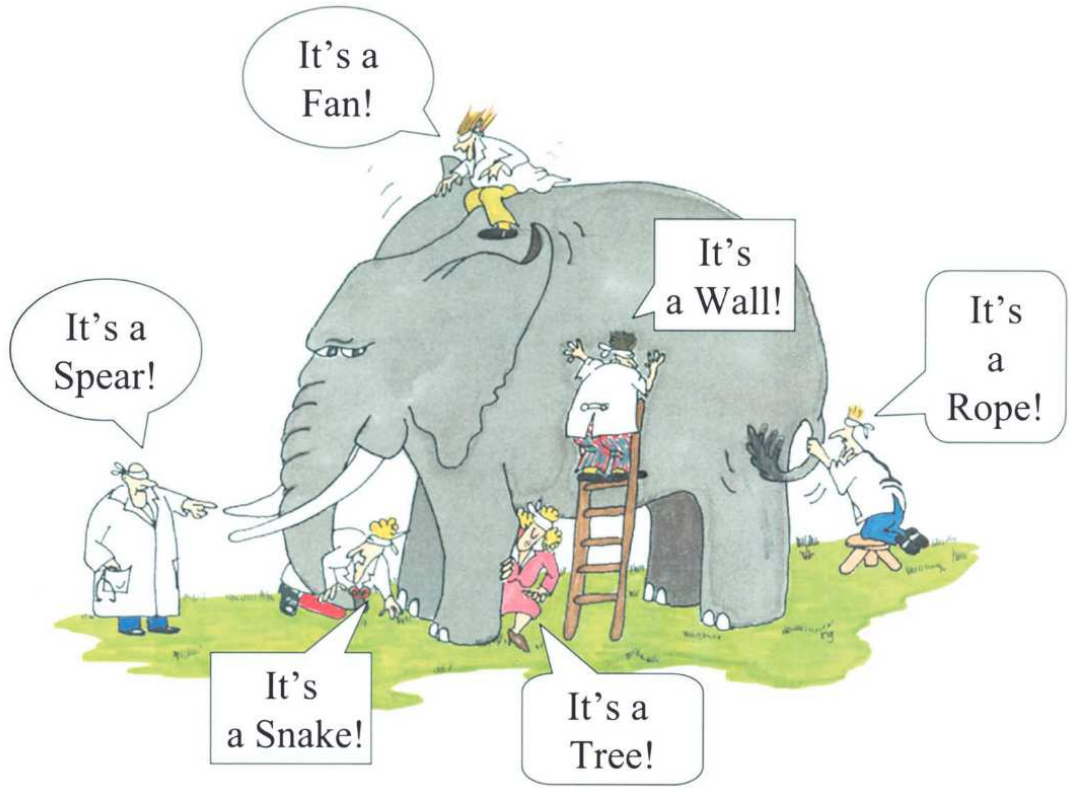}
\caption{
\textbf{The elephant and the six blind men.} (Cartoon originally copyrighted by the authors of \cite{himmelfarb_2002}; G. Renee Guzlas, artist). Taken from \cite{himmelfarb_2002}.
\label{elephant}
}% end of caption
\end{center}
\end{figure}

By analogy to the story of the blind men and the elephant (Fig.~\ref{elephant}), each scientist has a strong knowledge and supposedly lots of data on a particular area of the elephant, but no one has the knowledge that in reality what they are observing is an elephant. No one of these observations can provide a global view of the unifying concept. The story is well described in the poem \textit{The Blind Men and the Elephant} of the American poet John Godfrey Saxe (1816-1887) that can be found in \cite{himmelfarb_2002}.

Another interesting question related to understanding and prediction is the issue of understanding machines, which has been investigated by some researchers \cite{thorisson_2017,bieger_2017}, though apparently not so many. They consider that for the term "Understand" to be useful in the field of AI, it must refer to something measurable. Among the criteria to consider, they mention: (1) to predict the behavior of the phenomenon, (2) to achieve the objectives regarding the phenomenon, (3) to explain the phenomenon and (4) to create or recreate the phenomenon. In any case the notion of understanding by machines is by no means a simple problem.

\section{Machine Learning and scientific method}

Recent advances in machine learning and the associated hype behind has provoked the appearance of AI enthusiasts and skeptics. There are some enthusiasts of the AI that have dared to announce even the end of the scientific method as we know it today \cite{anderson_2008}. Other enthusiasts pretend to extract predictions and even natural laws by simply using experimental data \cite{schmidt_2009} or even creating machines for scientific discovery able to win a Nobel prize \cite{kitano_2016}. Not to mention the recent book by Max Tegmark on being human in the age of Artificial Intelligence \cite{tegmark_2017}. 

A few years ago, a provocative article published by Chris Anderson, editor in chief of the magazine Wired, with the title \textit{The End of Theory: The Data Deluge Makes the Scientific Method Obsolete} \cite{anderson_2008} provoked a large discussion among scientists. In his article Anderson argued that it was enough to establish correlations by using enough data that eventually could be analyzed without any need for models or hypothesis. Basically, by throwing the data into the huge computers was enough letting only the algorithms to find statistical patterns.

Others argue that in some occasions there is a trade-off where we can renounce understanding, since obviously is more complicated than simply compute something and make some quick predictions.

Gary Smith in his recent book \textit{The AI Delusion} \cite{smith_2018} encourages scepticism about artificial intelligence and the blind trust we put in it. In a certain sense, his book represents a response to the philosophy represented by the article of Anderson, because unfortunately too many people have been attracted by these claims. He expresses it by explicitly writing: ``Far too many intelligent and well-meaning people believe that number-crunching is enough. We do not need to understand the world. We do not need theories. It is enough to find patterns in data. Computers are really good at that, so we should turn our decision-making over to computers.”

In reality, he explains with numerous examples why we should not be intimidated into thinking that computers are infallible, that data-mining is knowledge discovery, and that black boxes should be trusted, emphasizing the importance of human reasoning as fundamentally different from artificial intelligence, which is why is needed more than ever.

In spite of the enthusiasts denying the scientific method, there are voices that oppose this viewpoint and mark the limits of machine prediction. Among them we can cite \cite{hosni_2018a,coveney_2016,buchanan_2019,crutchfield_2014}.

In particular in \cite{hosni_2018a} the authors critically assess the claim that bigger data leads to bigger predictions. They use analogies and ideas from atmospheric sciences and essentially conclude that a compromise between modelling and quantitative analysis is the best strategy for forecasting, as already anticipated long ago by Lewis Fry Richardson and John von Neumann, as pioneers in numerical weather prediction. They highlight that too many data do not make necessarily more accurate predictions, as is well known in weather forecasts. They also emphasize the important role played by the high dimension of systems with a high enough number of degrees of freedom versus the intrinsic role of chaos as a limiting factor to predictability in low dimensional systems. All this is nicely described in great detail in \cite{cecconi_2012} and in other recent and enlightening papers by Angelo Vulpiani and collaborators \cite{baldovin_2018,hosni_2018b,vulpiani_2020}. Similar ideas have been also recently defended by Mark Buchanan \cite{buchanan_2019} as well, arguing that the limits on the predictive accuracy of big data is derived from the theory of dynamical systems in the context of high-dimensional systems, the case in many typically complex problems like the weather and other real-world applications. This same idea was already commented when the new notion of hetero-chaos was discussed in Sect. 3.

Analogously Jim Cruthfield \cite{crutchfield_2014} argues in a fantastic manner on the importance of combining data, theory and computations, and intuition.

A defense of the scientific method versus the mere analysis of data is well documented in \cite{coveney_2016} in the context of biological and medical sciences. The authors clearly point out the weaknesses of pure big data approaches that cannot provide a true understanding and conceptual vision of the physical processes involved and subsequent applications. They make a strong defense of the theory as a guide to experimental design and to produce reliable predictive models and conceptual knowledge and understanding. They also remark the importance for biology and bioinformatics students to be trained to understand the theory of dynamical systems that are needed to describe and model biological dynamical processes.

Their skepticism brings them to affirm ``More attention needs to be given to theory if the many attempts at integrating computation, big data and experiment are to provide useful knowledge. A substantial portion of funding used to gather and process data should be diverted towards efforts to discern the laws of biology." And one of the authors rhetorically had expressed it with the following sentence: ``Does anyone really believe that data mining could produce the general theory of relativity?" \cite{dougherty_2011}. In another recent article \cite{succi_2019} Sauro Succi and Peter Coveney argue that some extravagant claims of big data need to be revised in view of some obstacles such as nonlinearity, non-locality, and high dimensionality derived from the science of complex systems, defending a hybrid method where data and theory come together, and somehow improving the scientific method as we know it.

\begin{figure}
\begin{center}
\includegraphics[width=0.7 \textwidth]{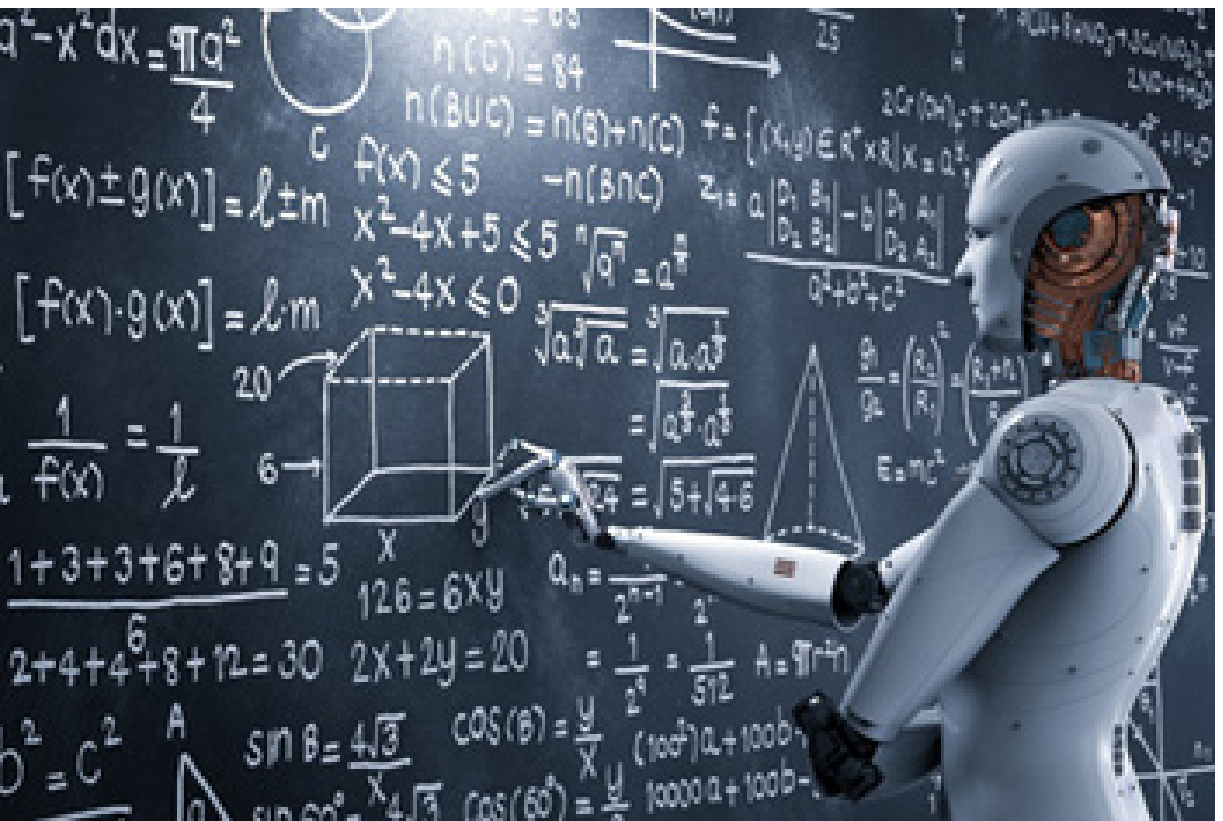}
\caption{The figure illustrates one of the dreams of AI, a robot attempting to do maths. We are very far from that.
\label{robot_math}
}% end of caption
\end{center}
\end{figure}

In an interesting document published by \textit{edge.org} and edited by John Brockman in 2015, a key question was asked to numerous scientists and artists including Nobel Prize winners about \textit{What do you think about machines that think?} \cite{edge_2015}. Nearly two hundred responses are included, where one can see all kind of responses, from enthusiasts to skeptics and between. I have selected here the response of a well-known physicist, Freeman Dyson, that is concise, surprising and with a bit of humour: "I could be wrong: I do not believe that machines that think exist, or that they are likely to exist in the foreseeable future. If I am wrong, as I often am, any thoughts I might have about the question are irrelevant. If I am right, then the whole question is irrelevant." Figure~\ref{robot_math} shows a machine doing mathematics.

In the context of geosciences and weather prediction is worth to mention here a fascinating recent book written by an atmospheric physicist, Shaun Lovejoy \cite{lovejoy_2019}, who besides leads the new discipline of nonlinear geophysics \cite{lovejoy_2009}. He strongly emphasizes the idea that concepts of nonlinear geophysics, mainly derived from nonlinear dynamics, fractal geometry and complex systems theory, can provide a rational basis for the statistics and models of natural systems, making our understanding of the world more complete. Furthermore, he makes a detailed discussion on the limits of predictability either by using the standard deterministic chaotic models or the lesser known stochastic models in weather predictions.

Of great interest on our discussion on prediction and understanding are the insightful comments on the current role played by theory and quick numerical results in atmospheric science, and how this is affecting understanding. He writes: "Theory of any kind was increasingly seen as superfluous; it was either irrelevant or a luxury that could no longer be afforded. Any and all atmospheric questions were answered using the now- standard tools: NWPs and GCMs. Unfortunately, these models are massive constructs built by teams of scientists spanning generations. They were already “black boxes,” and even when they answered questions, they did not deliver understanding. Atmospheric science was gradually being transformed from an effort at comprehending the atmosphere to one of imitating it numerically (i.e., into a purely applied field). New areas— such as the climate— were being totally driven by applications and technology: climate change and computers. In this brave new world, few felt the need or had the resources to tackle basic scientific problems."

Likewise in the context of geosciences, a nice perspective article was recently published in Nature \cite{reichstein_2019} where the authors defend similar ideas, focusing mainly in geoscientific data, and analysing the relationship between deep learning and process understanding in data-driven Earth system science. They review in a superb manner the developments of machine learning in geosciences, and discussed that there are certain predictive problems related to forecasting extreme events such as floods or fires or predicting in the biosphere, where not substantial advances have been seen in the past few years, in spite of the deluge of data that we are accumulating nowadays. In few words, there has not been much progress in prediction even though the capacity to accumulate more data has tremendously increased. 

They unreservedly defend that the most promising and challenging future would be to gain understanding in addition to optimizing prediction, so that they propose an integration of machine learning with physical modelling. The idea is that data-driven machine learning approaches will successfully complement and enrich the physical modelling, featuring a conceptualized and interpretable understanding. Precisely one of the challenges they establish for deep learning methods is the need for understanding and for what they call interpretability, and causal discovery from observational data. Definitely, machine learning methods provide an excellent improvement of classification and prediction, but it does not help much to scientific understanding.

\section{Conclusions}

We are witnessing an era in which big data and machine learning and deep learning techniques will contribute, as they are already doing, in a very important way to the advancement of science, whether applied or basic. Numerous examples in many different disciplines have illustrated the powerful predictive capabilities of these techniques, including examples in chaotic dynamics. All this has created an enormous hype on the new possibilities, and further creating very high expectations for the future. Likewise, in the face of some perhaps exaggerated positions about the potential of these techniques, a reaction has been provoked in the scientific community by pointing out the flaws of these positions, as well as some limits, sometimes affecting the core of the scientific method.

As a result of these efforts, it can be concluded that we cannot do without the role of modeling, conceptualization and other tools provided by theoretical science and scientific method, when one of the important goals is understanding. Prediction and understanding are two fundamental ideas that should guide us about the progress of science.

I want to emphasize again here the importance of the dynamical systems for the process of understanding and to get insights about the physical and biological processes involved in our observations and describe and model them.

There is no doubt that the path of the future of science will be marked by a constructive dialogue between big data and big theory. Data science has much to contribute, but without theoretical and conceptual models we cannot understand.

Despite all the above, there are some who think that one day the machines will be able to carry out all the activities that the human brain is capable of doing. If we extend it to scientific creation, as well as to the possibility of finding new laws of physics and to the same elaboration of scientific theories, we could conclude that man's contribution to science would have ended. No matter how much excitement the machine learning techniques are creating, it seems that we are very far from that and, therefore, we have as humans much future ahead to discover, understand and predict.

\begin{acknowledgments}
I acknowledge an interesting encounter and a further discussion with Mark Barthelemy, after which he encouraged me to write this article. Some of these ideas were exposed in a meeting around the question "Is it possible that Artificial Intelligence generate scientific theories?" that took place at the Spanish Royal Academy of Sciences for what I wish to acknowledge José María Fuster and Jesús M. Sanz-Serna. In addition, I thank Angelo Vulpiani and Peter Coveney for providing me some new interesting references. This work was supported by the Spanish State Research Agency (AEI) and the European Regional Development Fund (ERDF) under Project No. FIS2016-76883-P.
\end{acknowledgments}

\section*{Conflict of interest} The author declares that he has no conflict of interest.

\end{document}